

%
\newbox\leftpage \newdimen\fullhsize \newdimen\hstitle \newdimen\hsbody
\tolerance=1000\hfuzz=2pt
\def\bigans{b }
\message{ big or little (b/l)? }\read-1 to\answ
\ifx\answ\bigans\message{(This will come out unreduced.}
\magnification=1200\baselineskip=24pt plus 2pt minus 1pt
\hsbody=\hsize \hstitle=\hsize 
\else\def\apans{l }\message{ lyman or hepl (l/h) (lowercase!) ? }
\read-1 to \apansw\message{(This will be reduced.}
\let\lr=L
\magnification=1000\baselineskip=16pt plus 2pt minus 1pt
\voffset=-.31truein\vsize=7truein
\hstitle=8truein\hsbody=4.75truein\fullhsize=10truein\hsize=\hsbody
\ifx\apansw\apans\special{ps: landscape}\hoffset=-.59truein
  \else\hoffset=.05truein\fi
\output={\ifnum\pageno=0 
  \shipout\vbox{\hbox to \fullhsize{\hfill\pagebody\hfill}}\advancepageno
  \else
  \almostshipout{\leftline{\vbox{\pagebody\makefootline}}}\advancepageno
  \fi}
\def\almostshipout#1{\if L\lr \count1=1
      \global\setbox\leftpage=#1 \global\let\lr=R
  \else \count1=2
    \shipout\vbox{\ifx\apansw\apans\special{ps: landscape}\fi 
      \hbox to\fullhsize{\box\leftpage\hfil#1}}  \global\let\lr=L\fi}
\fi
%
\catcode`\@=11 
\newcount\yearltd\yearltd=\year\advance\yearltd by -1900

\def\Title#1#2{\nopagenumbers\abstractfont\hsize=\hstitle\rightline{#1}%
\vskip 1in\centerline{\titlefont #2}\abstractfont\vskip .5in\pageno=0}
\def\Date#1{\vfill\leftline{#1}\tenpoint\supereject\global\hsize=\hsbody%
\footline={\hss\tenrm\folio\hss}}
\def\draftmode{\def\draftdate{{\rm preliminary draft:
\number\month/\number\day/\number\yearltd\ \ \hourmin}}%
\headline={\hfil\draftdate}\writelabels\baselineskip=20pt plus 2pt minus 2pt
{\count255=\time\divide\count255 by 60 \xdef\hourmin{\number\count255}
	\multiply\count255 by-60\advance\count255 by\time
   \xdef\hourmin{\hourmin:\ifnum\count255<10 0\fi\the\count255}}}

\def\nolabels{\def\eqnlabel##1{}\def\eqlabel##1{}\def\reflabel##1{}}
\def\writelabels{\def\eqnlabel##1{%
{\escapechar=` \hfill\rlap{\hskip.09in\string##1}}}%
\def\eqlabel##1{{\escapechar=` \rlap{\hskip.09in\string##1}}}%
\def\reflabel##1{\noexpand\llap{\string\string\string##1\hskip.31in}}}
\nolabels
%
\global\newcount\secno \global\secno=0
\global\newcount\meqno \global\meqno=1
\def\newsec#1{\global\advance\secno by1
\xdef\secsym{\the\secno.}\global\meqno=1
\bigbreak\bigskip
\noindent{\bf\the\secno. #1}\par\nobreak\medskip\nobreak}
\xdef\secsym{}
\def\appendix#1#2{\global\meqno=1\xdef\secsym{\hbox{#1.}}\bigbreak\bigskip
\noindent{\bf Appendix #1. #2}\par\nobreak\medskip\nobreak}
%
%
\def\eqnn#1{\xdef #1{(\secsym\the\meqno)}%
\global\advance\meqno by1\eqnlabel#1}
\def\eqna#1{\xdef #1##1{\hbox{$(\secsym\the\meqno##1)$}}%
\global\advance\meqno by1\eqnlabel{#1$\{\}$}}
\def\eqn#1#2{\xdef #1{(\secsym\the\meqno)}\global\advance\meqno by1%
$$#2\eqno#1\eqlabel#1$$}
%
\newskip\footskip\footskip14pt plus 1pt minus 1pt 
\def\f@@t{\baselineskip\footskip\bgroup\aftergroup\@foot\let\next}
\setbox\strutbox=\hbox{\vrule height9.5pt depth4.5pt width0pt}
\global\newcount\ftno \global\ftno=0
\def\foot{\global\advance\ftno by1\footnote{$^{\the\ftno}$}}
%
%
\global\newcount\refno \global\refno=1
\newwrite\rfile
\def\ref{[\the\refno]\nref}
\def\nref#1{\xdef#1{[\the\refno]}\ifnum\refno=1\immediate
\openout\rfile=refs.tmp\fi\global\advance\refno by1\chardef\wfile=\rfile
\immediate\write\rfile{\noexpand\item{#1\ }\reflabel{#1}\pctsign}\findarg}
\def\findarg#1#{\begingroup\obeylines\newlinechar=`\^^M\pass@rg}
{\obeylines\gdef\pass@rg#1{\writ@line\relax #1^^M\hbox{}^^M}%
\gdef\writ@line#1^^M{\expandafter\toks0\expandafter{\striprel@x #1}%
\edef\next{\the\toks0}\ifx\next\em@rk\let\next=\endgroup\else\ifx\next\empty%
\else\immediate\write\wfile{\the\toks0}\fi\let\next=\writ@line\fi\next\relax}}
\def\striprel@x#1{} \def\em@rk{\hbox{}} {\catcode`\%=12\xdef\pctsign{
\def\semi{;\hfil\break}
\def\addref#1{\immediate\write\rfile{\noexpand\item{}#1}} 
\def\listrefs{\vfill\eject\immediate\closeout\rfile
\baselineskip=24pt\centerline{{\bf References}}\bigskip{\frenchspacing%
\escapechar=` \input refs.tmp\vfill\eject}\nonfrenchspacing}
\def\startrefs#1{\immediate\openout\rfile=refs.tmp\refno=#1}
\def\figures{\centerline{{\bf Figure Captions}}\medskip\parindent=40pt}
\def\fig#1#2{\medskip\item{Figure ~#1:  }#2}
\catcode`\@=12 
%
\ifx\answ\bigans
\font\titlerm=cmr10 scaled\magstep3 \font\titlerms=cmr7 scaled\magstep3
\font\titlermss=cmr5 scaled\magstep3 \font\titlei=cmmi10 scaled\magstep3
\font\titleis=cmmi7 scaled\magstep3 \font\titleiss=cmmi5 scaled\magstep3
\font\titlesy=cmsy10 scaled\magstep3 \font\titlesys=cmsy7 scaled\magstep3
\font\titlesyss=cmsy5 scaled\magstep3 \font\titleit=cmti10 scaled\magstep3
\else
\font\titlerm=cmr10 scaled\magstep4 \font\titlerms=cmr7 scaled\magstep4
\font\titlermss=cmr5 scaled\magstep4 \font\titlei=cmmi10 scaled\magstep4
\font\titleis=cmmi7 scaled\magstep4 \font\titleiss=cmmi5 scaled\magstep4
\font\titlesy=cmsy10 scaled\magstep4 \font\titlesys=cmsy7 scaled\magstep4
\font\titlesyss=cmsy5 scaled\magstep4 \font\titleit=cmti10 scaled\magstep4
\font\absrm=cmr10 scaled\magstep1 \font\absrms=cmr7 scaled\magstep1
\font\absrmss=cmr5 scaled\magstep1 \font\absi=cmmi10 scaled\magstep1
\font\absis=cmmi7 scaled\magstep1 \font\absiss=cmmi5 scaled\magstep1
\font\abssy=cmsy10 scaled\magstep1 \font\abssys=cmsy7 scaled\magstep1
\font\abssyss=cmsy5 scaled\magstep1 \font\absbf=cmbx10 scaled\magstep1
\skewchar\absi='177 \skewchar\absis='177 \skewchar\absiss='177
\skewchar\abssy='60 \skewchar\abssys='60 \skewchar\abssyss='60
\fi
\skewchar\titlei='177 \skewchar\titleis='177 \skewchar\titleiss='177
\skewchar\titlesy='60 \skewchar\titlesys='60 \skewchar\titlesyss='60
\def\titlefont{\def\rm{\fam0\titlerm}
\textfont0=\titlerm \scriptfont0=\titlerms \scriptscriptfont0=\titlermss
\textfont1=\titlei \scriptfont1=\titleis \scriptscriptfont1=\titleiss
\textfont2=\titlesy \scriptfont2=\titlesys \scriptscriptfont2=\titlesyss
\textfont\itfam=\titleit \def\it{\fam\itfam\titleit} \rm}
\ifx\answ\bigans\def\abstractfont{\tenpoint}\else
\def\abstractfont{\def\rm{\fam0\absrm}
\textfont0=\absrm \scriptfont0=\absrms \scriptscriptfont0=\absrmss
\textfont1=\absi \scriptfont1=\absis \scriptscriptfont1=\absiss
\textfont2=\abssy \scriptfont2=\abssys \scriptscriptfont2=\abssyss
\textfont\itfam=\bigit \def\it{\fam\itfam\bigit}
\textfont\bffam=\absbf \def\bf{\fam\bffam\absbf} \rm} \fi
\def\tenpoint{\def\rm{\fam0\tenrm}
\textfont0=\tenrm \scriptfont0=\sevenrm \scriptscriptfont0=\fiverm
\textfont1=\teni  \scriptfont1=\seveni  \scriptscriptfont1=\fivei
\textfont2=\tensy \scriptfont2=\sevensy \scriptscriptfont2=\fivesy
\textfont\itfam=\tenit \def\it{\fam\itfam\tenit}
\textfont\bffam=\tenbf \def\bf{\fam\bffam\tenbf} \rm}
%
%
\def\noblackbox{\overfullrule=0pt}
\hyphenation{anom-aly anom-alies coun-ter-term coun-ter-terms}
\def\inv{^{\raise.15ex\hbox{${\scriptscriptstyle -}$}\kern-.05em 1}}
\def\dup{^{\vphantom{1}}}
\def\Dsl{\,\raise.15ex\hbox{/}\mkern-13.5mu D} 
\def\dsl{\raise.15ex\hbox{/}\kern-.57em\partial}
\def\del{\partial}
\def\Psl{\dsl}
\def\tr{{\rm tr}} \def\Tr{{\rm Tr}}
\font\bigit=cmti10 scaled \magstep1
\def\biglie{\hbox{\bigit\$}} 
\def\lspace{\ifx\answ\bigans{}\else\qquad\fi}
\def\lbspace{\ifx\answ\bigans{}\else\hskip-.2in\fi} 
\def\boxeqn#1{\vcenter{\vbox{\hrule\hbox{\vrule\kern3pt\vbox{\kern3pt
	\hbox{${\displaystyle #1}$}\kern3pt}\kern3pt\vrule}\hrule}}}
\def\mbox#1#2{\vcenter{\hrule \hbox{\vrule height#2in
		\kern#1in \vrule} \hrule}}  
%
\def\CAG{{\cal A/\cal G}}   
\def\CA{{\cal A}} \def\CC{{\cal C}} \def\CF{{\cal F}} \def\CG{{\cal G}}
\def\CL{{\cal L}} \def\CH{{\cal H}} \def\CI{{\cal I}} \def\CU{{\cal U}}
\def\CB{{\cal B}} \def\CR{{\cal R}} \def\CD{{\cal D}} \def\CT{{\cal T}}
\def\e#1{{\rm e}^{^{\textstyle#1}}}
\def\grad#1{\,\nabla\!_{{#1}}\,}
\def\gradgrad#1#2{\,\nabla\!_{{#1}}\nabla\!_{{#2}}\,}
\def\ph{\varphi}
\def\psibar{\overline\psi}
\def\om#1#2{\omega^{#1}{}_{#2}}
\def\vev#1{\langle #1 \rangle}
\def\lform{\hbox{$\sqcup$}\llap{\hbox{$\sqcap$}}}
\def\darr#1{\raise1.5ex\hbox{$\leftrightarrow$}\mkern-16.5mu #1}
\def\lie{\hbox{\it\$}} 
\def\ha{{1\over2}}
\def\half{{\textstyle{1\over2}}} 
\def\roughly#1{\raise.3ex\hbox{$#1$\kern-.75em\lower1ex\hbox{$\sim$}}}

\Title{PUPT-91-1437 (1993), hep-ph/9401352}
{\vbox{
\centerline{Efficient Electroweak Baryogenesis from Lepton Transport} }}
\baselineskip 20pt
\centerline{\bf Michael Joyce}
\centerline{\bf Tomislav  Prokopec}
\centerline{and}
\centerline{\bf Neil Turok}
\centerline{Joseph Henry Laboratories }
\centerline{Princeton University}
\centerline{Princeton, NJ 08544}
\vskip .3in
\centerline{\bf Abstract}
\baselineskip 16pt
One mechanism for generating a baryon asymmetry at the
electroweak phase transition involves propagation of
particle asymmetries generated by reflection from
the bubble walls into the unbroken phase.
Hitherto attention has focussed on top quarks because
of their large mass and thus effective scattering
from bubble walls. In this paper we point out that leptons may
be more efficient mediators of this type of
electroweak baryogenesis, particularly for thicker bubble walls
favored by perturbative calculations.
We carry out an analytic calculation of each stage of the mechanism
and conclude it produces a baryon asymmetry
compatible with observations for a wide range of parameters.

\Date{October 1994 (Revised) }

\eject
\baselineskip 24pt
An exciting development over the last few years has been the
realization that the
standard electroweak theory, possibly with a minimal
extension of the Higgs sector, has the potential to produce the observed
matter-antimatter asymmetry in the universe.
The mechanism involves the remarkable anomaly
structure of the Weinberg-Salam theory, which ties changes in the topology
of the gauge and Higgs fields to changes in the baryon number of
the universe
\ref\KRS{
G. 't Hooft.
 Phys. Rev. Lett., {\bf 37} 8, 1976;
 A.D. Linde.
 Phys. Lett. {\bf 70B} 306, 1977;
V. Kuzmin, V. Rubakov and M. Shaposhnikov, Phys. Lett. {\bf 155B},
 36 (1985); F. Klinkhamer and N. Manton, Phys. Rev. {\bf D30}, 2212 (1984);
 P. Arnold and L. Mclerran, Phys. Rev. {\bf D37}, 1020 (1988).}.
It is an  intriguing notion
 that the baryon number violating processes were somehow
 biased at the electroweak phase transition, and subsequently suppressed,
 generating  a matter-antimatter  asymmetry rather naturally
 (for reviews see
\ref\review{N. Turok, in {\it Perspectives in Higgs Physics},
ed. G. Kane, pub. World Scientific, p. 300 (1992). },
\ref\cknrev{A. Cohen, D. Kaplan and A. Nelson, Preprint
 UCSD-PTH-93-02 (1993), to appear in Ann. Rev. Nucl. Part. Sci.}).

 Following
 suggestions of Shaposhnikov
 \ref\shap{M. E. Shaposhnikov, JETP Lett, {\bf 44}, 465 (1986); Nucl. Phys.
{\bf B287}, 757 (1987); Nucl. Phys. {\bf B299}, 797 (1988).}, and
McLerran \ref\lar{L. McLerran, Phys. Rev. Lett. {\bf 62}, 1075 (1989).},
Turok and Zadrozny showed that this actually happens
in minimal extensions of the standard model, with more than one
fundamental Higgs field. In these theories,
there is a $CP$ odd
Higgs field phase which changes in a definite manner
in bubble walls as the Higgs fields `roll'
from the unbroken to the broken symmetry `vacuum'. This
phase couples to the anomaly through a triangle diagram,
producing a biasing of the anomalous processes
\ref\NTJTZa{N. Turok and J. Zadrozny,
Phys. Rev. Lett. {\bf 65}, 2331 (1990).},
\ref\NTJTZb{N. Turok and J. Zadrozny,
Nuc. Phys. {\bf B 358}, 471 (1991).},
\ref\MSTV{L. McLerran, M. Shaposhnikov, N. Turok and M. Voloshin,
 Phys. Lett. { \bf  256B}, 451 (1991).}. As the bubble walls
 propagate through
 the universe, they leave behind a trail of baryons.

 Cohen, Kaplan and Nelson (CKN) proposed two potentially
more efficient mechanisms in the same theories, `spontaneous'
baryogenesis
\ref\cknplb{A. Cohen, D.  Kaplan and A.
 Nelson, Phys. Lett. {\bf B263}, 86  (1991).},
 and the `charge transport' mechanism
\ref\cknnpb{A. Cohen, D.  Kaplan and A.
 Nelson, Nuc. Phys. {\bf B373}, 453  (1992).},
\ref\cknplbb{A. Cohen, D. Kaplan and A.  Nelson, Phys. Lett. {\bf B294}
 (1992) 57. }.
The former was
argued to dominate in the case of thick bubble walls,
favored by perturbative calculations in the standard model
and the minimal two-Higgs extension
\ref\bub{ M. Dine, R. Leigh, P. Huet, A. Linde, and D. Linde,
Phys. Rev. D46, 550 (1992); B-H. Liu, L. Mclerran and N. Turok,
 Phys. Rev. {\bf D46}, 2668 (1992).}.

But the `charge transport' mechanism was nevertheless interesting
because for thin walls it could produce a very large asymmetry
as a result of its `non-locality' --
the CP violating effects on the wall produce an
asymmetry in left-handed top quark number which then propagates
far into the unbroken phase where the anomalous processes are unsuppressed.

In a companion paper \ref\JPTi{M. Joyce, T. Prokopec and
N. Turok, Princeton preprint PUP-TH-1436 (1993),
revised July 1994, to appear in Phys. Lett. {\bf B}.}
we have argued that CKN's
analysis of both mechanisms contains an important error.
Inappropriate constraints on {\it global\/} quantum numbers have been
imposed in chemical potential calculations in both cases.
Our conclusion is that `spontaneous' baryogenesis
does not occur as envisaged by CKN for typical
wall velocities and thicknesses. The effects of particle
transport must be included and new calculations are required
to establish how this mechanism then proceeds. However our
criticisms do not require such fundamental revision of
the `charge transport' mechanism and it is clear that it will
still result in efficient
baryogenesis. We have emphasised however that it is more
appropriate to consider the propagation into the unbroken phase
of the particle asymmetries which drive baryogenesis rather than
hypercharge as originally advocated by CKN, or a charge $X$
devised later to include the effects of screening \cknplbb.
The mechanism might more appropriately
be thought of as a `particle transport' mechanism.
In this paper we implement this mechanism in a new way. We
show that leptons may be efficient mediators of
this `nonlocal' electroweak baryogenesis,
and we show that this allows generation of a sufficiently large
baryon asymmetry in minimal
extensions of the standard model with
rather small $CP$ violating parameters,
even for the thicker walls favored by perturbative calculations.

The fact that leptons are only weakly coupled to the plasma
gives them several advantages over quarks in this mechanism. First it
allows us to compute the reflection asymmetry with reasonable accuracy
for a wide range of bubble wall widths. In contrast,
calculations of quark reflection asymmetries are plagued by strong thermal
effects, which one expects to wash out the phase coherence
effects which have been used to produce interesting CP violating
reflection.
Techniques to include this effect in calculations
have yet to be developed, but it seems unlikely that the
zero temperature  Dirac equation, even with one loop
real self-energy corrections, as in
\ref\FS{G. Farrar and M. Shaposhnikov, Phys. Rev. Lett.
{\bf  70}, 2833 (1993); CERN preprint CERN-TH.6732/93.}, is reliable.
Perturbative calculations
suggest the wall thickness is $ 2 m_H(T)^{-1} \approx 20 -40 T^{-1}$,
whereas the mean free time for strong scattering
processes \ref\BP{
E. Braaten and R. Pisarski, Phys. Rev. {\bf D 46}, 1829 (1992);
R. Pisarski, Phys. Rev. {\bf D 47}, 5589 (1993).}, \bub, is estimated
to be only a few thermal wavelengths.
Second, for thick bubble walls the asymmetry in top quark reflection
is small anyway, because, in order to show any asymmetry at all,
particles must have momenta $p_z >m(T)$, their thermal mass in the broken
phase. But if $p_z$ is larger than the inverse wall width,
the WKB approximation becomes good, and phase interference effects
are exponentially suppressed\footnote{$^\dagger$}{As we will discuss in
\ref\JPTii{M. Joyce, T. Prokopec and N. Turok, Princeton preprints
PUPT-1495  and PUPT-1496}
it is possible, even in the classical limit, to produce a
reflected chiral asymmetry. But even in this case the effects of strong
scattering must still be superimposed on the calculation.}.
Third, because leptons are not strongly interacting,
a lepton asymmetry diffuses much more
efficiently into the unbroken phase in front of the wall and
can bias anomalous processes for much longer than a quark asymmetry.
Finally, for the same reason, leptons do not partake in the
strong anomalous processes which tend to destroy  any axial
asymmetry in quarks and suppress baryon production
\ref\GS{G.F. Giudice and
M. Shaposhnikov, Phys. Lett. {\bf B326}, 118 (1994). }, \JPTii.

	At first sight leptons appear to suffer
one major deficiency as messengers
of $CP$ violation: lepton masses
are much smaller than  the top quark mass,
so leptons appear to be more weakly scattered by bubble walls.
This overlooks the fact that what is of importance is the
finite temperature tree-level mass. In theories
like the two-Higgs theory,
the ratio of Yukawa couplings
is {\it not} fixed by
the zero temperature masses. There is an extra
parameter -- the ratio of the {\it vev\/}s. To produce the
zero temperature mass hierarchy, one can tune  either the Yukawa
couplings, or the {\it vev\/}s, or both.
Indeed, `universal' Yukawa couplings are quite natural
in family symmetry schemes \ref\BS{ For a discussion of
phenomenological constraints on such theories, including
the possibility of observing enhanced flavor changing neutral currents
 in $b$ physics, see I.I. Bigi and A.I. Sanda and N.G. Uraltsev in \review.
See also M. Joyce and N. Turok, Nuc. Phys. {\bf B416} (1994) 389
and references therein.}.
But the finite temperature {\it vev\/}s will in
general not obey this tuning,
and thus there is no reason
why the tau lepton mass (for example)
need be much smaller than the top quark
mass at temperatures close to the electroweak transition.

To summarize: for  walls thick in comparison to the mean free path
of quarks, top quark reflection is
difficult to calculate reliably and, we expect,
exponentially small if the reflection arises from
quantum interference effects.
Leptons on the other hand, which only interact weakly with the plasma,
should be much more accurately described by these calculations
and in addition are transported very  efficiently and do not
see strong anomalous processes. For thin walls,
both quark and lepton reflection may be large, but the extra efficiency
of lepton transport again means that lepton mediated baryogenesis
can dominate, provided the tau Yukawa coupling is not much smaller
than the top Yukawa coupling.

We shall describe our calculations in three parts: a)  the determination of
the asymmetry in the reflected chiral flux,
 b)  the calculation of the diffusion of a chiral lepton
 asymmetry
 far in front of the bubble walls, and c) integration
 of the $B$ violating rate equation to determine the final
matter-antimatter  asymmetry.

$CP$ violating effects in
the reflection of quarks from bubble walls have been studied
in \cknnpb\ and \FS. The particles incident on the wall see a space dependent
complex mass, and the problem is simply to solve the Dirac equation
in this background. Calculating perturbatively in
${m_l / m_H}$ the leading term in the asymmetry is
\eqn\scat{\eqalign{
{\cal R}_{R\rightarrow L} - {\cal R}_{L\rightarrow R}
 \approx 2 \int_{-\infty}^\infty dz
{\rm Im} [m_l(z)m_l^*(\infty)] { 1 \over |p_z| }
 \equiv 2\theta_{CP} {|m_l|^2 \over m_H |p_z|}\, ,
 \qquad |m_l|<|p_z|<m_H \, , }}
where ${\cal R}_{R\rightarrow L}$ and ${\cal R}_{L\rightarrow R}$ are
the probability that a right-handed lepton with momentum
$p_z$ orthogonal to the wall reflects into a left-handed lepton,
and vice versa. The
restriction on $|p_z|$ arises because for $|p_z|< |m_l|\equiv |m_l(\infty)|$
all incident leptons are reflected, while
for $|p_z| > m_H$ the WKB approximation becomes good.
Note that the nonlocality of \scat\ is a result of
quantum mechanical interference effects.

We assume for simplicity that the wall velocity
is nonrelativistic.
Integrating \scat\ over the phase space densities of
particles incident on the wall, assumed to be moving
through the plasma with speed $v_w$, and accounting for
transmitted as well as reflected particles,
one finds the net flux of left and right-handed leptons
injected back into the unbroken phase is, to leading order in
${m_l / m_H}$, ${m_H / T}$ and $v_w$:
\eqn\scatb{\eqalign{
J^L(0)  = -J^R(0)
 \approx { {v_w   |m_l|^2 m_H \theta_{CP} } \over {4\pi^2} }
\equiv J_0\, .
}}
{}From $CPT$ invariance it follows
that exactly  opposite asymmetries are transmitted into the
broken symmetry phase \cknnpb.

We now wish to determine the effect of this injected chiral
asymmetry on $B$ violating processes.
First we need the rate equation for baryon production
in the unbroken phase
in the presence of a nonzero chemical potential for
the participating fermions (see e.g. \JPTi):
\eqn\rat{\eqalign{
\dot{B} = - N_f { \Gamma_s \over T} (3 \mu_{q_L} + \mu_{l_L})\, ,
}}
where $N_f$ is the number of families, $\Gamma_s$ is the
rate for anomalous processes, $\Gamma_s =\kappa \alpha_W^4 T^4$,
where $\kappa$ is determined numerically
to be $\sim 0.1-1.0$
\ref\sph{ T. Askaard, H. Porter and
M.E. Shaposhnikov, Phys. Lett. {\bf 244B},
479 (1990); Nucl. Phys {\bf B353} 346 (1991)}
, and $\mu_{q_L}$ and $\mu_{l_L}$ are
the sum of the chemical potentials for left-handed quarks and
left-handed leptons, respectively.
For the case when one neglects top quark reflection,
$3\mu_{q_l}+\mu_{l_L}=\mu_L=(6/T^2) L_L$, where $L_L$ is the number
density of left-handed leptons. The problem is now simply to
determine how these chemical potentials are driven by the injected
flux in front of the wall.

Let us briefly state and justify the assumptions which we make
in this calculation.
We assume  that the departure from thermal equilibrium
which is caused by the injected flux is modelled by a single
function for each particle species,
the space-time dependent chemical potential.
The injected flux (as exemplified by \scat\ ) has however
in general a very non-trivial momentum dependence. The assumption
should be justified nevertheless if the typical injected particle
spends long enough in the unbroken phase to thermalize.
To quantify this suppose that $\tau$ is the mean time for
a particle's velocity to be randomized.
This $\tau$ can be taken to be both a thermalization time and
the step time in an isotropic random walk which the particle executes
once it thermalizes. Then we can estimate the mean distance a particle
moves away while diffusing
in the direction of the wall motion in the plasma frame
in time $t$ to be
$ \sqrt { { t\tau \over 3} }$.
Equating this distance
with $v_wt$ to determine the time the particle actually spends in
the unbroken phase, we find that the ratio of the `injection time'
to `diffusion time' is $\sim 3v_w^2$. For the
non-relativistic wall velocities we will consider
it should thus be a good approximation to take the reflected particles
to spend most of their time
in the unbroken phase in a  diffusing thermalized flux.
We note that the approximation should be good
provided $v_w<v_s={1 \over \sqrt{3}}$, the speed of sound in the
relativistic plasma.

For a single species such a locally thermalized flux would be described
in general by a chemical potential, a temperature and a velocity. We
drop the latter two parameters because the processes
which equalize the temperature and velocity of different species
locally  are much faster than
those which damp the chemical potentials. We also neglect the gradients
in temperature and velocity assuming these to be efficiently
damped  for the same reason.

So we take the system to be described by a chemical potential (or
number density) for
each species which is spatially varying (in the rest frame of the wall).
The injected flux will source the equations describing these
chemical potentials, and the originally sourced species (in our
case the left- and right-handed leptons) will feed into other
species through various processes e.g. a lepton changes chirality by
scattering off a $W$ boson and emits a Higgs particle.
For the purposes of this letter we will neglect the
chemical potentials generated in species other than those
directly sourced by the flux. In a longer publication
\JPTii\
we will include these chemical potentials ( e.g. for Higgs particles)
and show that they bring about minor numerical changes to the results
derived here.

Let us emphasize that these assumptions are quite different
to those made in the previous literature.
In their calculation CKN characterized the injected flux in terms
of an injected charge ( originally
hypercharge, later a global $X$ charge related to it)
and calculated a local thermal equilibrium in the
presence of a chemical potential for this charge,
imposing constraints on all the exactly conserved global charges
such as $B-L$. As argued in \JPTi\ these charges should not be
fixed locally but should be determined dynamically. This is
precisely what we are doing in the present calculation - our
solution will correspond to a constrained local thermal
equilibrium, but with constraints which are self-consistently
determined by the solution of equations subject to the correct
boundary conditions. In the case of lepton
reflection it is clear for example that the induced local
$B - L$ is nonzero (even though it is zero
in the injected flux) because
of the different propagation properties of the left and
right-handed leptons.

Finally we remark on the role of screening. Khlebnikov
\ref\Khleb{S. Khlebnikov, Phys. Lett. {\bf B300}, 376(1993).}
pointed out that the injected flux in this mechanism carries
non-zero hypercharge, which will be screened. This effect can be included
by imposing  perfect screening as a constraint at each point in space,
as suggested by
CKN in {\cknplbb}. This describes the screening of a static potential
correctly, but not the dynamical case of a potential moving with
respect to the plasma. To model this properly one should add
a force term to the diffusion equations discussed below which
dynamically enforces $Y\approx 0$. We relegate this complication to
{\JPTii}, where we find that the injected current induces a
cancelling hypercharge current dominated by the particles
with  transport properties similar  to those of the injected particles.
Because there are three families
 to screen the
injected chiral asymmetry in the third family, the corrections to
the results below are minor numerical ones.

The dominant scattering processes for the leptons are
those mediated by weak gauge bosons. We have calculated the
diffusion constant due to such processes by solving
the relativistic Boltzmann equation
\JPTii.  Following the standard procedure \ref\Lif{E.M.
Lifshitz and L.P. Pitaevskii, {\it Physical Kinetics\/},
Pergamon Press (1979).} one introduces a chemical potential
$\mu(z)$ assumed to vary slowly in space. One then seeks a stationary
solution to the Boltzmann equation in which the deviation from the
zeroth order phase space density is of order $\partial_z \mu$.
The collision integral is computed in the approximation that
the energies of the scattered and incident particles are equal
Finally the flux induced by $\partial_z \mu$ is computed and
from the relation $D \equiv - J_z/\partial_z n$ with $n$ the number
density, one reads off the diffusion constant $D$.
The diffusion treatment is justified provided that fractional
change in the number density over a diffusion length $D$
is slow.

We find that the diffusion constants $D_L$ and $D_R$ for right
and left-handed leptons respectively are given by
\eqn\diff{\eqalign{
D_L^{-1}\approx 8\alpha_W^2 (2+0.7tan^4 \theta_W) T \approx {T \over 110}
 \qquad \qquad D_R^{-1} \approx 23 \alpha_W^2 tan^4 \theta_W T \approx {T \over
470}
}}
where $\theta_W$ is the Weinberg angle. Note the
very large diffusion length for right-handed leptons:
$D_R \approx 470/T \approx 4 D_L$.

The wall does not alter the total number of particles; rather
it produces a surplus of left-handed particles on one side,
and a deficit on the other side. This means there is a `blip' in
the current $J_z$ for left-handed particles, and a `blip'
of the opposite sign for right-handed particles.
In view of the arguments we have made we describe this as
\eqn\inj{\eqalign{
J^L(z) &\approx  \xi^L J_0\delta  (z-vt)   \qquad
J^R(z) \approx  -\xi^R J_0\delta(z-vt)  \cr }}
where $J_0$ is given in \scatb, and $\xi^{L,R}$ define the persistence
lengths of the left and right-handed currents in the vicinity
of the wall.

These parameters $\xi^{L,R}$ parametrize our ignorance of the details
of the thermalization. A simple estimate of their magnitude is
as follows.  For a diffusing particle,  the velocity randomization
time is
$\tau \sim 6D$. Using this
as a decay time for a flux injected at velocity $v_i$,
we estimate $ \xi\sim  6D v_i$.
Thus $\xi / D \sim 6v_i \approx 2 {m_H \over T}$,
where we have computed the mean velocity $v_i$
of the particles injected from the wall.
Our calculation of the reflection of
particles from the wall into
the medium is only valid if
the mean free path is significantly larger than the
wall thickness. In fact, since most reflected particles have large
$p_\perp$, they typically traverse much more than a single wall thickness
in the process of reflecting off the wall.
In order to ignore interactions on the wall, one
should satisfy
$\tau_{sc} v_i>L$, where $\tau_{sc}$ is
the mean free time
and $v_i\sim 1/LT$.
Hence $LT<(\tau_{sc} T)^{1/2}\sim (1/\alpha)^{1/2}$, with
$\alpha$ the appropriate coupling.
This condition is unlikely to be satisfied
for quarks, where the free particle approximation undoubtedly overestimates
the injected chiral current, and even for leptons
we expect some suppression\footnote{$^\dagger$}{We thank
Larry McLerran for a discussion of  this point. See also
\ref\cline{J. Cline, preprint UMN-TH-1259-94, May 1994}. }.
In  this letter we shall for simplicity assume
that the wall is thin enough that particle interactions on the
wall do not suppress the injected chiral current.

The total particle current is the sum of the current induced by the wall,
Eq. \inj,   and the
diffusion term given by $-D \partial_z n$; the
relevant diffusion equation then
follows from conservation of particle number.
Next, we include interactions through which
left  and right-handed particles can
decay into one another through processes like those depicted
in Fig. 1. We have computed these diagrams and find for that the
rates per particle
\eqn\inj{\eqalign{
\Gamma_{L R} \approx 0.3 \alpha_W y_{\tau}^2 T\, , }}
with $y_{\tau}$ the lepton Yukawa coupling.

Finally, we make the assumption that the densities $L_L$ and $L_R$
settle down to a stationary state, in which they are functions of $z-vt$.
The diffusion equations in the wall frame then become:
\eqn\diff{\eqalign{
D_L L_L'' + v_w L_L' - \Gamma_{LR} ({1\over2}L_L - L_R) &=
 \xi^L J_0
\delta'(z) \cr
D_R L_R'' + v_w L_R' + \Gamma_{LR} ({1\over2}L_L - L_R) &=
- \xi^R J_0
\delta'(z) \cr }}
with primes denoting $\partial_z$, the spatial derivative in the wall frame.
To a good approximation we can  ignore the
anomalous processes \rat\ in \diff\  -- as long as
the final baryon asymmetry is small, there is little feedback from it.

Solving these equations is a straightforward
matching problem, once one notices
that the equations possess an exactly conserved quantity obtained by
adding them, and integrating once:
\eqn\leptonnumber{
D_L L_L^\prime +D_R L_R^\prime +v_w (L_L+L_R)
}
This is  simply  lepton number conservation.

The parameter range we consider in detail here is
\eqn\param{\eqalign{
v_w^2 << \Gamma_{LR} D_R <1 .
}}
In this regime we see the interesting
effects of the efficient diffusion of right-handed leptons,
and at the same time our treatment of the slow decay of the
chiral asymmetry in the diffusion approximation \diff\
is well justified. The dominant baryon asymmetry comes from
a long diffuse tail of right-handed leptons. In this tail,
the left-handed leptons are maintained in an abundance $L_L
\approx 2 L_R$ dictated by the $\Gamma_{LR}$ processes. The
distributions of left and right-handed leptons in the tail are given by
\eqn\soln{\eqalign{ L_R(z) \approx {1\over 2} L_L(z) \approx
    -{J_0\over 2} {\xi^L \over D_L} e^{-\lambda_s z} \, ,   }}
where $\lambda_s \approx 3v_w/(2D_R+D_L)$
is the smallest  eigenvalue. A simple discussion following Eq. \inj\
implies that ${\xi^L / D_L}\approx {\xi^R / D_R }\sim 6v_i$.

Using \soln\
we can then integrate \rat\ in front of the bubble wall
to determine the baryon asymmetry the wall leaves behind it.
Putting everything together, we find for the baryon-to-entropy ratio
\eqn\ass{\eqalign{
{n_B \over s}\approx 0.04 \theta_{CP}{\kappa \alpha_W^2 cot^4 \theta_W \over
   g_* v_w}({m_{\tau}\over T})^2 {m_H \over T}{\xi_L \over D_L}   \, , }}
where the entropy density  $s=(2\pi^2/45)g_* T^3$ and   $g_* \approx 10^2$
is the number of relativistic degrees of freedom at the weak
scale.
 With our estimate $\xi^L/D_L \sim 2 m_H/T$, and using
$v_w \sim 0.1$,  we have ${n_B / s} \approx
10^{-6} \kappa \theta_{CP} y_{\tau}^2$. With $y_\tau \sim 0.1$
we find that the  baryon asymmetry is $ \sim 10^{-8} \theta_{CP}$.
For a standard model tau lepton Yukawa coupling
$y_\tau \sim 10^{-2}$ the condition \param\ requires
$v_w \le 0.01$ in which case the asymmetry is $\sim 10^{-9} \theta_{CP}$.
We shall discuss the result for other values of $y_\tau$ and $v_w$
in \JPTii.
In particular, note that the approximation made in (10)
of ignoring the anomalous processes breaks down at very low wall velocities.
In the case of a very slow wall,
$v_w<(\Gamma_s D_R)^{1/2}\sim 0.01$,
an equilibrium involving the anomalous processes
is reached in the right-handed tail, so that
$B_L\approx L_L/3$. As $v_w$ is taken to zero,
the baryon asymmetry vanishes linearly in $v_w$.
This is easily understood, since the injected
chiral flux provides an upper bound on the final
asymmetry, and this flux is proportional to $v_w$.

How efficient are tau leptons as compared to top quarks? As we pointed out,
the calculation is substantially  more difficult for the top quark case.
A crude estimate indicates that the ratio of
the baryon asymmetry coming from reflected top quarks to that from
leptons
is $\sim (\alpha_W/\alpha_S)^2 \tan^4(\theta_W) (y_t/y_\tau)^2 \sim
10^{-2} (y_t/y_\tau)^2$. This neglects any suppression due
to strong anomalous processes which, as we will discuss in \JPTii,
can take various forms depending on the wall velocity.
In any case, for thin walls, and `standard model'
like Yukawa couplings, top quark baryogenesis may dominate
in certain regimes.
The case of thicker walls for top quarks remains open,
requiring new calculations \JPTii.

In conclusion, we have attempted to carefully investigate
a new mechanism through which $CP$ violation present
in bubble walls at the electroweak phase transition can
result in
a bias of the anomalous processes.
It seems likely
that this mechanism is the dominant one in  electroweak
baryogenesis schemes for the expected case of `medium thick' walls,
and  a baryon  asymmetry
of $\sim 10^{-8} \times$ $CP $
violating phases may be produced.

\centerline{Acknowledgements}
We thank P.J.E. Peebles and M. Shaposhnikov for discussions.
The work of T.P. and N.T. is partially	supported by
NSF contract PHY90-21984, and the David and Lucile Packard
Foundation. M.J. is supported by a Charlotte Elizabeth Procter Fellowship.

\listrefs

\figures
\fig{1}{ Diagrams through which a left-handed lepton is changed
into a right-handed one, and vice versa.}
\bye